\documentclass[draft,grl]{agutex}

 \usepackage{lineno}
 \linenumbers*[1]

  \usepackage[dvips]{graphicx}
  \setkeys{Gin}{draft=false}
\authorrunninghead{Garc\'ia Mu\~noz et al.}

\titlerunninghead{Volcanic aerosols in a lunar eclipse}

\begin{document}

%% ------------------------------------------------------------------------ %%
%
%  TITLE
%
%% ------------------------------------------------------------------------ %%

\title{The impact of the Kasatochi eruption on the Moon's illumination 
during the August 2008 lunar eclipse}

%\title{%The somewhat diffuse origin of the eclipsed Moon's color\\

%
% e.g., \title{Terrestrial ring current:
% Origin, formation, and decay $\alpha\beta\Gamma\Delta$}
% You may use \\ to break the title over several lines.

%% ------------------------------------------------------------------------ %%
%
%  AUTHORS AND AFFILIATIONS
%
%% ------------------------------------------------------------------------ %%

\authors{A. Garc\'ia Mu\~noz,\altaffilmark{1,2}
       E. Pall\'e,\altaffilmark{1,2}
       M.R. Zapatero Osorio,\altaffilmark{3}
       and E.L. Mart\'in \altaffilmark{3}}

%Use \author{\altaffilmark{}} and \altaffiltext{}

% \altaffilmark will produce footnote;
% matching altaffiltext will appear at bottom of page.
% May use \\ to start a new line.

%\authors{R. C. Bales, \altaffilmark{1}
%E. Mosley-Thompson, \altaffilmark{2} R. Williams, \altaffilmark{3}
%and J. R. McConnell\altaffilmark{4}}

 \altaffiltext{1}{Instituto de Astrof\'isica de Canarias (IAC), C/ V\'ia L\'actea s/n,
 E-38200 La Laguna, Tenerife, Spain} 
 \altaffiltext{2}{Departamento de Astrof\'isica, 
 Universidad de La Laguna (ULL), E-38206 La Laguna, Tenerife, Spain} 
 \altaffiltext{3}{Centro de Astrobiolog\'ia, CSIC-INTA, Madrid, Spain} 

%\altaffiltext{1}{Department of Hydrology and Water Resources,
%University of Arizona, Tucson, Arizona, USA.}

%\altaffiltext{2}{Department of Geography, Ohio State University,
%Columbus, Ohio, USA.}

%\altaffiltext{3}{Department of Space Sciences, University of
%Michigan, Ann Arbor, Michigan, USA.}

%\altaffiltext{4}{Division of Hydrologic Sciences, Desert Research
%Institute, Reno, Nevada, USA.}

%% ------------------------------------------------------------------------ %%
%
%  ABSTRACT
%
%% ------------------------------------------------------------------------ %%

% >> Do NOT include any \begin...\end commands within
% >> the body of the abstract.
 
\begin{abstract} 
The Moon's changeable aspect during a lunar eclipse is largely
attributable to variations in the refracted unscattered
sunlight absorbed by the terrestrial atmosphere that occur as the satellite 
crosses the Earth's shadow.
%Also, 
The contribution to the Moon's aspect %during a lunar 
from sunlight scattered at the Earth's terminator 
is generally deemed minor.
However, our %model-based 
analysis of a published spectrum of the 16 August 2008 
 lunar eclipse shows that diffuse sunlight is a major 
component of the measured spectrum at wavelengths shorter than 600 nm. 
The conclusion is supported by two distinct features, namely the spectrum's
tail at short wavelengths
and the unequal absorption by an oxygen collisional complex 
at two nearby bands. 
Our findings are consistent with the presence of the volcanic cloud 
reported at high northern latitudes 
following the 7--8 August 2008 eruption  in Alaska of the Kasatochi volcano. 
The cloud both 
attenuates the unscattered sunlight and enhances 
%\textbf{
moderately
%} 
the scattered component,
thus modifying the contrast between the two contributions. 

\end{abstract}

%% ------------------------------------------------------------------------ %%
%
%  BEGIN ARTICLE
%
%% ------------------------------------------------------------------------ %%

% The body of the article must start with a \begin{article} command
%
% \end{article} must follow the references section, before the figures
%  and tables.

\begin{article}

%% ------------------------------------------------------------------------ %%
%
%  TEXT
%
%% ------------------------------------------------------------------------ %%

\section{Introduction}
%A lunar eclipse occurs when the Earth intercepts the view of the 
%solar disk from the Moon. In the eclipse,  
%some sunlight permanently reaches the Moon because the
%sunbeam is refracted in the %dense layers of the 
%terrestrial atmosphere. 
%It is generally accepted that 
%both brightness and color of the eclipsed Moon are primarily 
%dictated by the differential absorption of unscattered, 
%or direct, sunlight that occurs in the Earth's atmosphere. 
%Arguably, 
%lunar eclipses
%%the eclipsed Moon's changing color 
%provide some of the
%most recognizable images of an astrophysical event. % to non-specialists.

The classical theory of lunar eclipses  
accounts for refraction, differential absorption and focusing 
to explain the Moon's aspect %photometric images 
during an eclipse \citep{link1962}. 
Link's classical theory has been subsequently perfected and used to 
investigate the composition and aerosol loading of the 
Earth's atmosphere \citep[e.g.,][]{ugolnikovmaslov2008}. %Indeed, 
Aerosols play a critical role in the interpretation
of lunar eclipses as their content, distribution and optical properties are
largely unpredictable. 
Volcanic eruptions and meteor showers are two 
natural sources of aerosols with the potential for perturbing the atmosphere and, in
turn, the aspect of the eclipsed Moon \citep{keen1983,vollmergedzelman2008}. 
%\textbf{
Occasionally, large wildfires may also perturb the atmosphere 
\citep{frommetal2010}.
%}

\citet{garciamunozpalle2011} have revisited the lunar eclipse theory
and estimated the impact of volcanic aerosols on the spectrum of sunlight
at the eclipsed Moon.
%estimated the diffuse component 
%in spectra of the sunlight that arrives at the eclipsed Moon. 
%In elevated amounts, 
Aerosols may substantially 
attenuate 
the direct sunlight while simultaneously enhancing 
%\textbf{
somewhat
%} 
the 
scattered contribution. 
The latter depends strongly on the  capacity of aerosols 
for forward-scattering the incident light and, consequently, on the
aerosols' size. 
The spectroscopic characterization of the sunlight reflected from the eclipsed Moon
takes the investigation of lunar eclipses farther 
than allowed for by photometry and the traditional color indices.

The  7--8 August 2008 eruption of the Kasatochi volcano 
(52.17$^{\circ}$N, 175.51$^{\circ}$W, Aleutian Islands, Alaska) 
ended a period %, started in 1997, 
%characterized by 
of 
%\textbf{
global
%} 
low stratospheric  aerosol amounts.
%That period was preceded by some years
%of dilution of the %volcanic 
%matter injected 
%by the 1991 eruption of Pinatubo.
The three main explosions recorded %at Kasatochi 
over two days
%event,
plus the %continuous 
release of gas that followed 
for %nearly 16 
hours delivered into the atmosphere 
%about 
$\sim$1.5 Tg of SO$_2$ \citep{waythomasetal2010},
which is $\sim$30
 times less than the SO$_2$ injected by Pinatubo in 1991.
The plume of gas and ash 
%reached the stratosphere 
%that formed above the volcano 
rose 
%crossed the 
%tropopause
up to $\sim$14--18 km and 
%The volcanic cloud %of gas and ash 
drifted eastward carried by jet winds, %at near-tropopause altitudes, 
 spreading rapidly over North America, Greenland, and 
%a large expanse of 
the North Atlantic Ocean. 
The cloud was spotted above Europe on 15 August, 
one week after the eruption \citep{martinssonetal2009}. 
%The progress of the volcanic cloud was monitored 
%with unprecedented detail %for a volcanic event 
%from satellites, aerial platforms and the ground. 

Our paper shows that a published spectrum of the Moon in umbra during 
the August 2008 lunar eclipse 
contains sunlight scattered at the Earth's terminator.
% at short wavelengths.
We argue that the Kasatochi eruption is the most plausible origin for the 
abnormally elevated atmospheric opacity needed to 
explain the observation.
\citet{vidalmadjaretal2010} have published a spectrum of the
August 2008 lunar eclipse, but covering only the penumbra.

\section{Data} 
 
We use data of the 16 August 2008 lunar eclipse 
obtained with the ALFOSC instrument mounted on the
Nordic Optical Telescope at the Observatorio del Roque de los Muchachos 
(La Palma, Spain) and presented by \citet{palleetal2009}.
The dataset comprises spectra of the Moon 
in umbra (21:36UT), 
penumbra (22:11UT), and 
out of eclipse (23:09UT). 
%Airmass stands for the secant of the 
%telescope's pointing zenith angle.
\citet{palleetal2009} derived a lunar eclipse spectrum 
from the ratio of umbra and penumbra spectra. 
%If %the two spectra are 
%obtained at comparable airmasses, 
The ratio %practically 
cancels out
the solar spectrum and the telluric signature of the Moon-to-telescope 
optical path.
What remains is the imprint of the limb-viewed terrestrial atmosphere 
(averaged in a particular way over the terminator) 
on the sunlight that reaches the Moon in umbra. 
%\citet{palleetal2011} investigate the usefulness of lunar eclipse 
%spectra in the characterization of Earth-like exoplanets' atmospheres. 
%The characterization of the atmospheres of this particularly
%interesting family of exoplanets, though, 
%will have to await the construction of the coming generation of extremely large 
%telescopes. 
Our analysis sets out from the 400--900 nm published spectrum.

The solar elevation angle, $e$, is 
the geocentric angle between the incident sunbeam
direction and the direction from the Earth's centre to 
the lunar disk parcel targeted by the telescope. 
We have that $e$$\sim$0.34$^{\circ}$ for the slit projected on the Moon.
The structure of an umbra spectrum is very sensitive to $e$
\citep{garciamunozpalle2011}. 

%Figure (\ref{raypathmap_fig}) displays the great circle 
%that joins the subsolar point and the subslit point at 21:36UT on the date
%of the eclipse.
%The arc traces the projected mid-section of the sunbeam that
%reaches the tracked lunar disk. 
%Before arriving at the Moon, the sunbeam 
%crossed the bulk of the volcanic cloud over the North Atlantic Ocean.

\section{Evidence of scattered sunlight}
%This section discusses the two arguments that prove the occurrence of scattered
%sunlight in the measured spectrum.

%Two arguments prove the occurrence of scattered
%sunlight in the measured spectrum.

\subsection{The short-wavelength tail of the spectrum}  

Model predictions for $e$$\sim$0.34$^{\circ}$ and 
a broad range of aerosol loadings show that the spectrum of sunlight directly
transmitted through the atmosphere is typically 
2--3 orders of magnitude fainter at 400 nm than at 600 nm 
\citep{garciamunozpalle2011}.
This is at odds with the eclipse data, 
which show that the measured spectrum is roughly flat to within a 
factor of 2 shortwards of 600 nm and that 
the fluxes at 400 and 880 nm are in a  ratio of $\sim$1:20. 
%As a consequence, 
It thus means that direct sunlight is not the only contributor to the
measured spectrum. 
\citet{garciamunozpalle2011} note that a flat 
spectrum at short wavelengths indicates that diffuse sunlight dominates locally
over direct sunlight. %This is further explored below.

\subsection{The (O$_2$)$_2$ bands at 577 and 630 nm}
We fitted synthetic curves of the form $\prod_i\exp{(-\tau_i)}$ 
to the measured spectrum from 550 to 660 nm. 
The curves include %$\tau_i$'s for 
absorption by H$_2$O, O$_3$, O$_2$ and the (O$_2$)$_2$ collisional complex. 
%Including NO$_2$ and NO$_3$ was deemed inconsequential so they were omitted. 
One term, 
$\tau_{\rm{cont}}$=$\sum_{k=0}^{4} c_k (\lambda_*/\lambda)^k$, 
with $\lambda_*$= 600 nm, 
accounts for a continuum baseline. % by 
%Rayleigh scattering and aerosols. 
%In $\tau_{\rm{cont}}$, 
%We take . 
Thus, each curve contains up to ten degrees of freedom, 
namely, five $c_{k}$'s,
integrated columns for H$_2$O, O$_3$ and O$_2$, 
and, optionally, one integrated column for each of the 
[$X$$^3\Sigma_g^-$(0)]$_2$$\rightarrow$$a$$^1\Delta_g$(0)+$a$$^1\Delta_g$(1)
and
[$X$$^3\Sigma_g^-$(0)]$_2$$\rightarrow$[$a$$^1\Delta_g$(0)]$_2$
bands of (O$_2$)$_2$ 
%$X$(0)+$X$(0)$\rightarrow$$a$(0)+$a$(1) and 
%$\rightarrow$$a$(0)+$a$(0) bands of (O$_2$)$_2$ 
that occur at 577 and 630 nm, respectively.
For the  temperature-dependent gas properties,  
the temperature was fixed at 225 K. 
The synthetic curves were properly degraded and resampled. %accordingly.
%to the resolution of the experiment 
%of the observation %, $\sim$1000, 
%and resampled. % at the wavelengths of the measurements.
The minimization of %the 
%summed 
%residual 
$\chi^2$=$\sum_{j} 
(1-I_{\rm{fit}}(\lambda_j)/I_{\rm{exp}}(\lambda_j))^2$,
%\textbf{
where $I_{\rm{fit}}$ and $I_{\rm{exp}}$ are the synthetic and observed
data,
%}  
outputs the best fit parameters. 
 
\begin{figure}
\noindent\includegraphics[width=20pc]{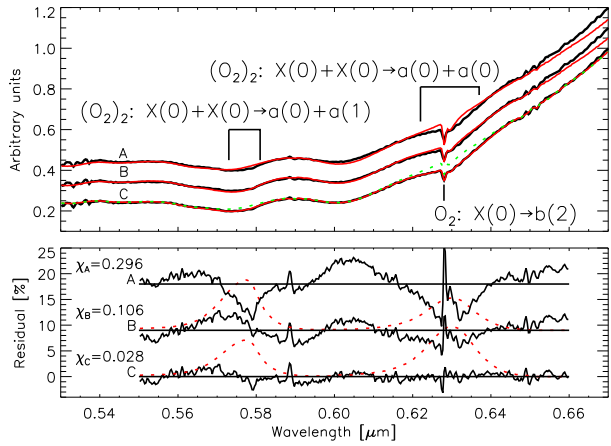}
 \caption{\label{fit1_fig} Top: Best fits (red) to the measured 
 spectrum (black) from 550 to 660 nm. 
 %Fit A includes neither of the 577 or 630
 %bands of (O$_2$)$_2$; 
 %B includes both bands and assumes a common integrated column; 
 %C includes both bands and allows for separate
 %integrated columns for each of them. 
 The dotted green line are the best fits
 divided by $\exp{(-\tau_{(\rm{O}_2)_2})}$.
The comparison of the dotted and solid curves 
makes explicit the contributions from the coincidental in position, 
albeit distinct in nature, 
{O$_2$ $X$(0)$\rightarrow$$b$(2)} and 
{(O$_2$)$_2$ $X$(0)+$X$(0)$\rightarrow$$a$(0)+$a$(0)} absorption bands 
near 630 nm. 
Bottom: Fit residuals. For B and C, the dotted red curves are
the (O$_2$)$_2$ contributions. In C, we infer an optimal ratio for the
577:630 nm integrated columns of $\sim$1:2.}
\end{figure}

Figure (\ref{fit1_fig}) summarizes the best fits obtained
from three separate strategies, each of them treating  
the 577 and 630 nm  bands of (O$_2$)$_2$ in a different manner.
Strategy A fits the spectrum with null amounts of (O$_2$)$_2$; 
B includes (O$_2$)$_2$ and assumes the same integrated column
for the two bands; and %, 
%finally, 
C allows for separate integrated columns 
for each of the 577 and 630 nm bands. 
In the top panel, the  solid black curves represent the measured 
spectrum, shifted in the vertical for comparison with the synthetic curves. 
The red solid curves are the respective A, B and C best fits. 
%In addition, for C, the graph shows in green dots the best fit after division by 
%$\exp{(-\tau_{(\rm{O}_2)_2})}$. 
The bottom panel displays the residuals. % for each of the A, B and C 
%best fits. 
Including (O$_2$)$_2$ reduces notably the fit residuals. 
The fit improves further if the integrated column at 630 nm is 
about twice the column at 577 nm. The latter conclusion is the core of the
second argument that proves the significance of diffuse sunlight in the
measured lunar eclipse spectrum. Some comments on the robustness of the
fitting procedure can be found in the Supplementary Material.

%The unequal integrated columns inferred for the (O$_2$)$_2$ bands 
Taking C as the optimal strategy, 
the conclusion 
is that average sunlight photons at 577 and 630 nm follow different paths
in the atmosphere.
The direct trajectories of sunlight rays are dictated by 
the atmospheric refractive index, which does not change appreciably 
within such a narrow spectral interval. 
The amount of sunlight directly transmitted does however 
vary sharply with wavelength. 
Direct sunlight is more attenuated at 577 nm than at 630 nm
due to the $\sim$$\lambda^{-4}$ behaviour of the Rayleigh cross section 
and the closer proximity of the 577 nm band to the absorption peak of 
the O$_3$ Chappuis band. 

%We thus have to invoke the sunlight scattered in the terrestrial atmosphere to
%explain the observation.
We thus have to invoke 
sunlight scattered at the Earth's terminator to explain the measured spectrum.
%the contribution of
%scattered sunlight. %in the terrestrial atmosphere.
%Scattered sunlight photons are redirected moonward from 
%a broad range of altitudes at the Earth's terminator. The actual 
%range depends on the photon wavelength and both the
%amount and distribution of aerosols.  
\citet{garciamunozpalle2011} show that in a lunar eclipse 
the bulk of diffuse sunlight near 600 nm originates from above 15 km. 
In the stratosphere, the (O$_2$)$_2$ density, 
which drops 
%\textbf{
with a scale height half that of the background density, 
%}
 is negligibly small. 
Foreseeably, 
the %absorption 
signature of the (O$_2$)$_2$ bands in the diffuse sunlight spectrum 
 is weak.
  
\section{Analysis and discussion}

Next, we generate model lunar eclipse spectra 
that reproduce the measured spectrum 
if a few reasonable assumptions on the
loading and properties of airborne aerosols are introduced. 
The spectra,
generated with the model described by \citet{garciamunozpalle2011},
contain both direct and diffuse components. 
Further details on the %background atmosphere and 
underlying model assumptions 
can be found in the Supplementary Material.

\begin{figure}
\noindent\includegraphics[width=20pc]{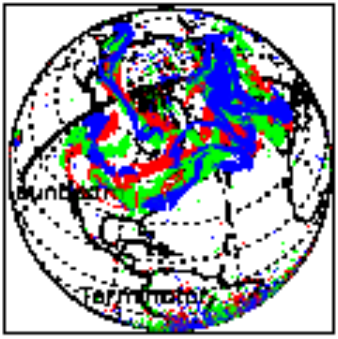}
\caption{\label{raypath_fig} 
Solid: Projected mid-section trajectory 
 of the sunbeam that reaches the lunar 
disk targeted by the telescope at 21:36UT on 16 August 2008. 
Overplotted, the SO$_2$ cloud (a usual volcanic cloud tracer) on 
15, 16 and 17 August (red, green and blue, respectively) according to 
AURA/OMI data (downloaded from the Giovanni online data system, 
developed and maintained by the NASA GES DISC). 
The sunbeam's closest approach to the Earth's surface occurs in the North 
Atlantic region. 
%\textbf{
The local tropopause is at $\sim$10 km.
%} 
 Dashed: Line of the terminator. 
} 
\end{figure}

The tracing of the direct sunbeam 
that reached the parcel of the Moon's disk %in umbra 
tracked 
by the telescope reveals that the sunbeam 
intercepted the volcanic cloud formed in the
Kasatochi eruption, as seen in Fig. (\ref{raypath_fig}). 
%Figure (??) %in the Supplementary Material 
%shows the projected sunbeam trajectory 
%on the planet 
%and the cloud location on 15--17 August.
It is expected that 
the direct sunlight component is more strongly affected by the volcanic cloud
than the diffuse one, which 
originates from all terminator locations.
%including regions far from the cloud. 
This distinction is accounted for in the generation of the model spectra by
assuming separate
atmospheres for the calculation of each component. 
%We take the 1.02-$\mu$m extinction profiles retrieved from 
%Atmospheric Chemistry Experiment Imager
%solar occultation data  in September 2007 and 2008 
%as our reference aerosol extinction profiles \citep{siorisetal2010}.

For simplicity, the model spectra are allowed only
four adjustable parameters. These are 
$f_{\gamma_{\rm{0}}}$, $\alpha'$ and $f_{\rm{O}_3}$ for the direct sunlight
component, and $r_{\rm{eff}}$ for the diffuse one. 
In the former, 
$f_{\gamma_{\rm{0}}}$ scales the reference aerosol extinction profile at 1.02
$\mu$m, $\alpha'$ is the {\AA}ngstr\"om exponent to extrapolate the 1.02-$\mu$m
extinction profile to shorter wavelenghts, and 
$f_{\rm{O}_3}$ scales the reference ozone profile. 
In the calculation of the diffuse sunlight component, 
$r_{\rm{eff}}$ stands for a mean effective radius for aerosols at
the terminator. 
%Typically, large $r_{\rm{eff}}$'s result in enhanced forward-scattering
%efficiencies.
%In the quiescent atmosphere, 
The sulfate droplets of background aerosols in the quiescent atmosphere have 
$r_{\rm{eff}}$$\sim$0.1--0.2 $\mu$m, whereas 
volcanic ash particles with residence times longer than a few days may have 
$r_{\rm{eff}}$'s of a few microns \citep{baumanetal2003b,munozetal2004}. 
\citet{siorisetal2010} report 
$r_{\rm{eff}}$'s of $\sim$0.6 $\mu$m for September 2008, which are indicative of a 
perturbed atmosphere. 
It is unclear what the mean $r_{\rm{eff}}$ at the terminator was 
one week after the eruption. 
Thus, we explored a set 
of $r_{\rm{eff}}$ from 0.1 to 2 $\mu$m to bracket possible sizes. 
A large $r_{\rm{eff}}$ results in phase functions strongly peaked in the forward
direction.

%A size of $r_{\rm{eff}}$$\sim$0.1 $\mu$m is representative of an 
%unperturbed atmosphere, whereas $r_{\rm{eff}}$$\sim$2 $\mu$m 
%would mean that the entire terminator is rich in fresh ash. 
%In going from %effective 
%radii of 0.1 to 2 $\mu$m, the phase function in the forward direction 
%augments by $\sim$20. 
%The phase functions are calculated at a few wavelengths and interpolated in 
%between. 
%Changing $r_{\rm{eff}}$ should be accompanied by changes in the 
%{\AA}ngstr\"om exponent utilized in the evaluation of the wavelength-dependent 
%extinction for the diffuse sunlight calculations. The many uncertainties 
%associated with the average properties of the atmosphere at the terminator 

\begin{figure} 
\noindent\includegraphics[width=39pc]{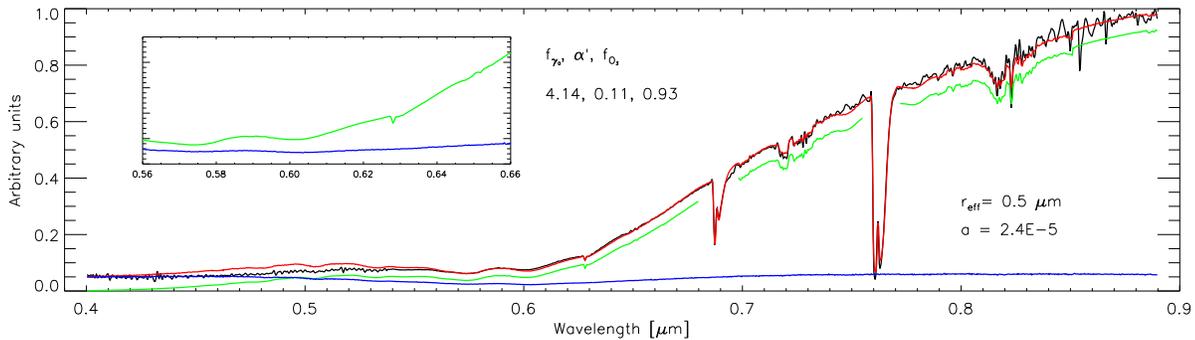} 
\caption{\label{fit2_fig} Best model fit for $r_{\rm{eff}}$=0.5 $\mu$m (red)
to the measured spectrum (black).
The model spectrum contains contributions from direct sunlight (green) and diffuse
sunlight (blue). The inset is a zoom of the region near 600 nm. The 
algorithm aims the fit of the continuum away from O$_2$ and H$_2$O bands.  
The  H$_2$O bands were fitted separately after the fit to the continuum. 
Figure (I) in 
the Supplementary Material shows the best fits for the full set of 
$r_{\rm{eff}}$ values.
}
\end{figure}

For each $r_{\rm{eff}}$, one diffuse sunlight spectrum was produced.
%In turn, 
For each diffuse spectrum 
an algorithm seeks the $f_{\gamma_{\rm{0}}}$, $\alpha'$ and $f_{\rm{O}_3}$ 
values producing the best fit of the direct + diffuse model spectra
to the continuum of the measured spectrum.
The algorithm forces the (flux-uncalibrated) measured spectrum to match 
the model spectra at 875 nm. 
%The partly-saturated O$_2$ bands at 689 and 762 nm 
%and all H$_2$O bands were excluded from the fit. 
Figure (\ref{fit2_fig}) shows the best fit for $r_{\rm{eff}}$=0.5 $\mu$m 
and the values inferred for the other three adjustable parameters. 
The $a$ parameter is the multiplicative factor 
to pass from the normalization in the graph
to the Earth-to-Sun ratio as discussed by \citet{garciamunozpalle2011}. 
Figure (I) in the Supplementary Material shows the best fits for 
the full $r_{\rm{eff}}$ set.
It is apparent 
the good \textit{a posteriori} match of the O$_2$ bands in all cases. 
%somehow confirms the validity of our approach. 
The differences between the measured spectrum and the best fits
are of a few percent longwards of 600 nm, 
but of $\sim$50{\%} near 500 nm. This is a consequence of 
fitting the measured spectrum with models that contain
 a reduced number of adjustable parameters. 
The residuals longwards of 700 nm are mainly due to a known instrumental issue of
uncorrected fringing.
%(www.not.iac.es/instruments/alfosc). 

The  $f_{\gamma_{\rm{0}}}$ values inferred point to heavy aerosol loadings with 
peak extinctions of $\sim$10$^{-2}$ km$^{-1}$
in the atmosphere intercepted by the direct sunbeam.
%on the order of . 
Comparable extinctions were reported on global scales 
for a few months after the Pinatubo eruption \citep{baumanetal2003b}. 
Exponents $\alpha'$$\sim$0--0.15 are indicative of large-size
particles being carried in the volcanic cloud. 
The conversion from SO$_2$ to sulfate droplets has an e-folding time of 
20--50 days \citep{kristiansenetal2010}. 
It is unlikely that one week after the eruption is enough time for 
large sulfate droplets to form. Thus, the $\alpha'$ values inferred  
suggest that the volcanic cloud carried sizeable amounts of unsedimented 
ash. \citet{siorisetal2010} report small 
{\AA}ngstr\"om exponents of $\sim$0.5 in early September 2008.
%for data averaged over the northern hemisphere.

The inset of Fig. (\ref{fit2_fig}) shows the two component
spectra near 600 nm.
%in the vicinity of the (O$_2$)$_2$ bands. 
The diffuse spectrum is roughly flat and shows no evidence of (O$_2$)$_2$ 
absorption. When the direct and diffuse model spectra are added, 
the 577 nm band becomes more diluted than the 630 nm band, which
%longer-wavelength one, which
translates into an effective integrated column at 630 nm larger than at 577 nm. 
For the case in Fig. (\ref{fit2_fig}) the %577:630 
ratio is $\sim$1:1.4,
somewhat smaller than the $\sim$1:2 ratio inferred 
from the measured spectrum.
One may generally state that comparable amounts of direct and scattered sunlight 
near 600 nm lead to larger (O$_2$)$_2$ columns at the
longer-wavelength band.

Figure (I) in the Supplementary Material proves that good fits 
to the measured spectrum are possible for the full $r_{\rm{eff}}$
set.
This means that the measured spectrum accepts one quantitative 
interpretation for each $r_{\rm{eff}}$. 
In qualitative terms, though, the picture that we obtain is fairly consistent
and indicates that the direct sunbeam was substantially attenuated by
the volcanic cloud,
which leads to an 
enhanced contrast of the diffuse component.
For future efforts, we suggest that 
the flux calibration of the undivided spectra might help break the degeneracy.

%We suggest for future attempts the flux calibration of the undivided spectra as
%a way to break the degeneracy. Properly flux-calibrated spectra will eliminate
%the need for normalizing the lunar eclipse spectrum at each step of the fitting
%algorithm.

%The model spectra are meant to reproduce the continuum of the 
%measured spectrum away from molecular bands.

A comment is to be made regarding the Ring effect and the 
structure seen in the measured spectrum shortwards of 540 nm. 
The Ring effect 
refers to the smearing of solar Fraunhofer lines that occurs in the spectrum of
sunlight scattered in the atmosphere \citep{graingerring1962}. 
In the Earth's atmosphere, 
the Ring effect is due to rotational Raman scattering by  N$_2$ and
O$_2$  \citep{kattawaretal1981}. 
Raman scattering redistributes in wavelength part of the 
%causes the wavelength-redistribution of part of the
incident photons, the redistribution being more evident where the incident
solar spectrum shows the sharpest lines.
The ratio of scattered to unscattered sunlight spectra %obtained at mid-to-low resolutions 
reveals the Ring effect as 
a filling-in of the solar line cores. 
The detection of the Ring effect in the eclipse data would
mean a further confirmation of scattered sunlight. 
The measured lunar eclipse spectrum shows that 
ripples do occur in the Fraunhofer region. 
The sign of the structures
is however inverted with respect to what the Ring effect would produce. 
The inspection of the undivided umbra spectrum shows that the solar lines are
unexpectedly deep, probably due to the limitation in the subtraction of the sky
spectrum at these wavelengths, where the signal-to-noise ratio is the lowest.
Thus, the structure seen in the measured spectrum cannot be attributed to the
Ring effect.
Further, a few quantitative arguments allow us to deem as minor the 
impact of the Ring effect on the measured lunar eclipse spectrum. 
Following \citet{kattawaretal1981}, the filling-in 
for forward-scattered sunlight is
$\sim$2.5\% of the continuum Rayleigh-scattered by the gas.
In the conditions explored here 
the filling-in would be undetectably small because the 
sunlight scattered by the gas contributes less than a few percent 
to the net sunlight scattered  by gas and aerosols together. 
 
%\textbf{ 
Pyrocumulonimbus (pyroCbs) is a recently-coined term to designate convective 
activity triggered or sustained by wildfires \citep{frommetal2010}. 
In extreme events, pyroCbs 
inject smoke and biomass-burning particles into the troposphere and lower 
stratosphere and alter the global aerosol loading. 
PyroCbs may result in aerosol 
extinctions $\sim$10$^{-3}$--10$^{-2}$ km$^{-1}$ well above the 
tropopause, opacities that are often associated with volcanic clouds. 
It would be difficult to differentiate the impact 
on the eclipsed Moon of one such event from that of a volcanic eruption. 
To our knowledge, 
no extreme pyroCbs were reported in the weeks preceding the eclipse, 
a period that was monitored with unprecedented detail. 
Thus, if any, the contribution in the eclipse of pyroCbs 
blended with that of the Kasatochi cloud. 
%} 
 
% /VELO/GRL2011/COMPUTATIONS/S07/FEATURELESS

%Some words on the Perseid meteor shower that was taking place
%simultaneous to the eclipse.
%Episodic meteor showers deposit substantial amounts of dust in the atmosphere. 
%Micronic and smaller particles formed in the physico-chemical
%transformation of the entering meteoroids have residence times of days to weeks,
%temporarily altering the  optical atmospheric properties.  
The Perseids is one of the most copious meteor showers, 
running yearly from late July to late August.
In 2008, its peak of activity occurred near 13 August. 
%, a few days ahead of the lunar eclipse. 
Despite recent work
\citep{mateshvilietal1999,renardetal2010}, 
there are significant uncertainties 
on the optical properties of the atmosphere
perturbed by meteor showers. 
\citet{mateshvilietal1999} report two-fold enhancements
with respect to pre-shower values in the twilight 
brightness above 20 km during the Leonids in 1998. 
Assuming that both meteor showers are comparable and that the
brightness enhancement translates into a similar increase in  stratospheric 
opacity, the effect of extraterrestrial dust would be more than one
order of magnitude smaller than that by the volcanic perturbation. 
Thus, %it appears that 
the effect of meteoroid dust in the measured spectrum is likely masked by the 
volcanic perturbation. 

We have shown that the lunar eclipse spectrum published by 
\citet{palleetal2009} 
was affected by sunlight scattered at the Earth's terminator. 
We offered theoretical arguments that hint at 
the Kasatochi eruption as the most plausible origin for the 
atmospheric perturbation needed to explain the observations. 
Future observations will allow us to compare lunar eclipse spectra 
obtained in different atmospheric conditions. 
In a broader context, it is worth mentioning that 
the retrieval of globally-averaged atmospheric optical properties 
%for the Earth's atmosphere
is a relevant exercise towards the future characterization 
of transiting 
%\textbf{
Earth-like extrasolar planets.
%}
%This illustrates some of the possibilities offered by 
%lunar eclipse spectra in the investigation of the Earth's atmosphere. 
As a corollary, we may state that the color of the lunar
disk in umbra during the 16 August 2008 lunar eclipse was partly caused by 
diffuse sunlight.

\begin{acknowledgments}
ELM acknowledges a Visiting Research Professorship at the Department of Geological
Sciences of the University of Florida. We thank the two reviewers for
constructive comments.
%The calculation of the optical properties of aerosols is based on the Mie-theory
%code developed by M. Mishchenko and available at 
%http:$//$www.giss.nasa.gov$/$staff$/ \\
%/$mmishchenko$/$t$\_$matrix.html.
\end{acknowledgments}

%% ------------------------------------------------------------------------ %%
%
%  REFERENCE LIST AND TEXT CITATIONS
%
% Either type in your references using
% \begin{thebibliography}{}
% \bibitem{}
% Text
% \end{thebibliography}
%
% Or,
%
% If you use BiBTeX for your References, please produce your .bbl
% file and copy the contents into your paper here.
%
% Follow these steps:
% 1. Run LaTeX on your LaTeX file.
%
% 2. Run BiBTeX on your LaTeX file.
%
% 3. Open the new .bbl file containing the reference list and
%   copy all the contents into your LaTeX file here.
%
% 4. Comment out the old \bibliographystyle and \bibliography commands.
%
% 5. Run LaTeX on your new file before submitting.
%
% AGU does not want a .bib or a .bbl file, but asks that you
% copy in the contents of your .bbl file here.

%Reference citation examples:

%...as shown by \textit{Kilby} [2008].
%...has been shown [\textit{Kilby et al.}, 2008].

%...as shown by \cite{jskilby}.
%...has been shown \citep{jskilbye}.

%% ------------------------------------------------------------------------ %%
%
%  END ARTICLE
%
%% ------------------------------------------------------------------------ %%

\end{article}

%% Enter Figures and Tables here:

% When submitting articles through the GEMS system:
% COMMENT OUT ANY COMMANDS THAT INCLUDE GRAPHICS.

\newpage

\clearpage

%\begin{figure}
%\centering
%\includegraphics[width=15cm]{sup1.eps}
%\end{figure}

\begin{figure}
\centering
{\bf Supplementary material}
\includegraphics[width=15cm]{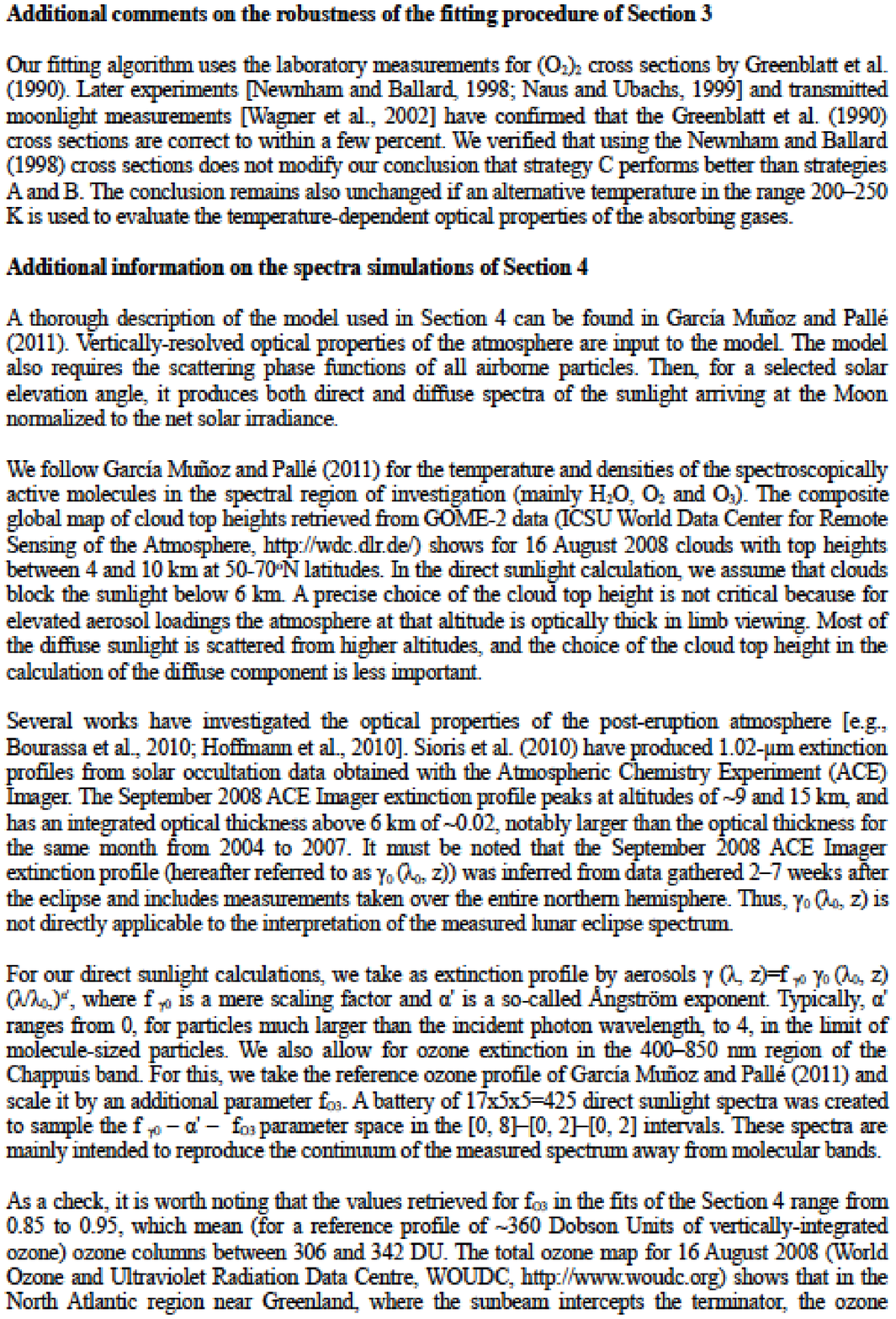}
\end{figure}

\begin{figure}
\centering
\includegraphics[width=15cm]{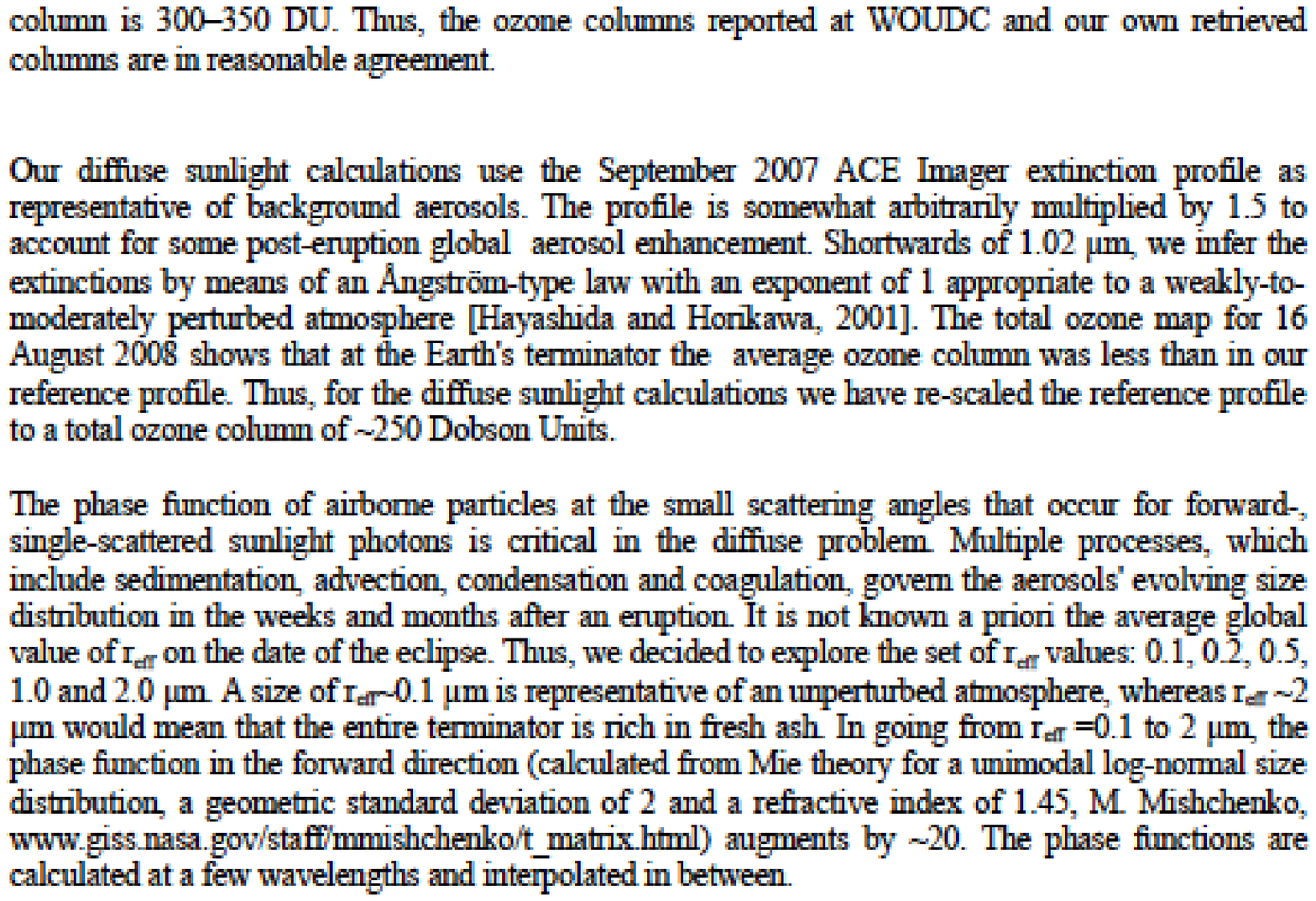}
\end{figure}

\begin{figure}
\centering
\includegraphics[width=15cm]{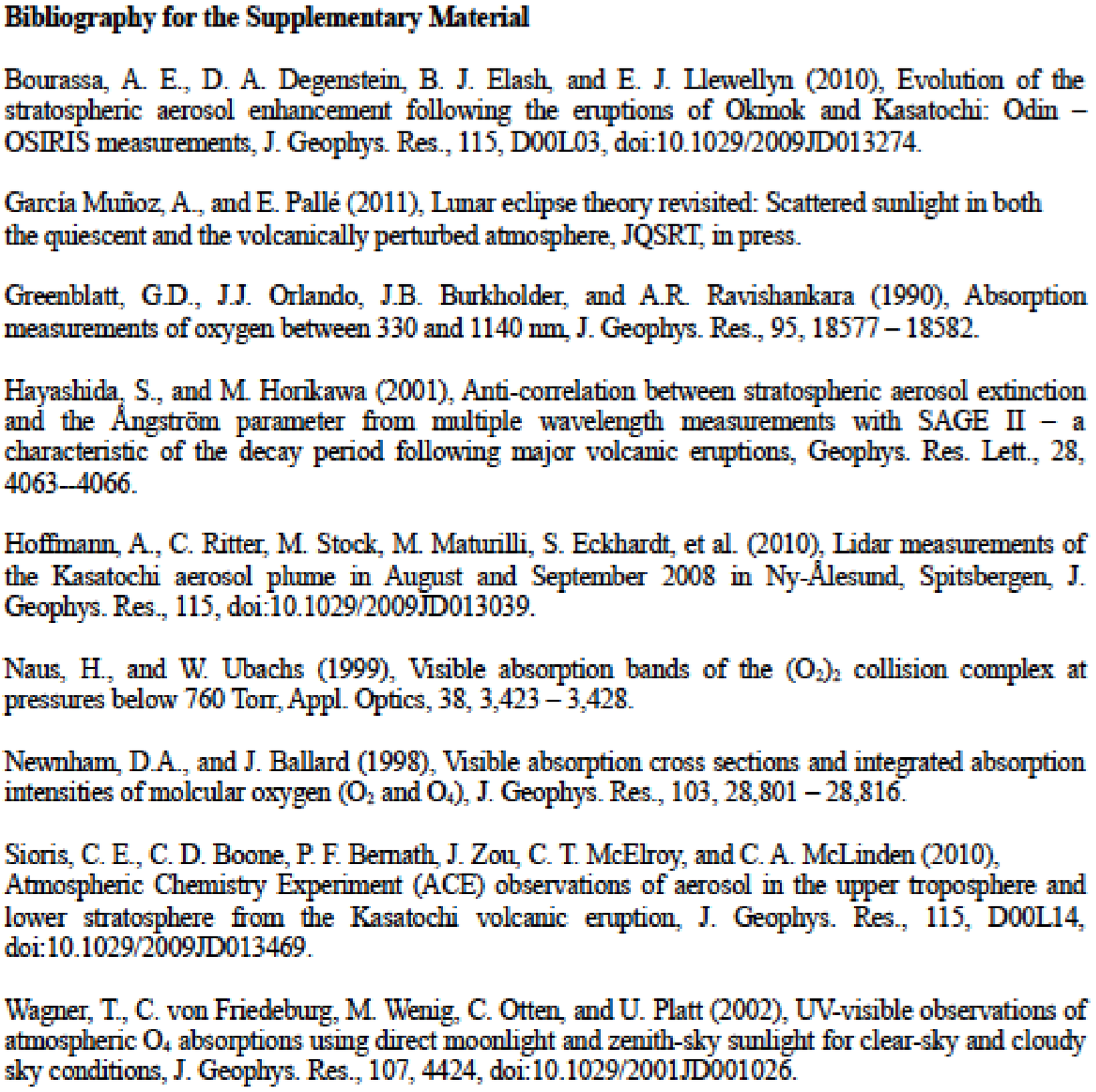}
\end{figure}

\begin{figure}
\centering
Figure mentioned in the main text. \\
\includegraphics[width=15cm]{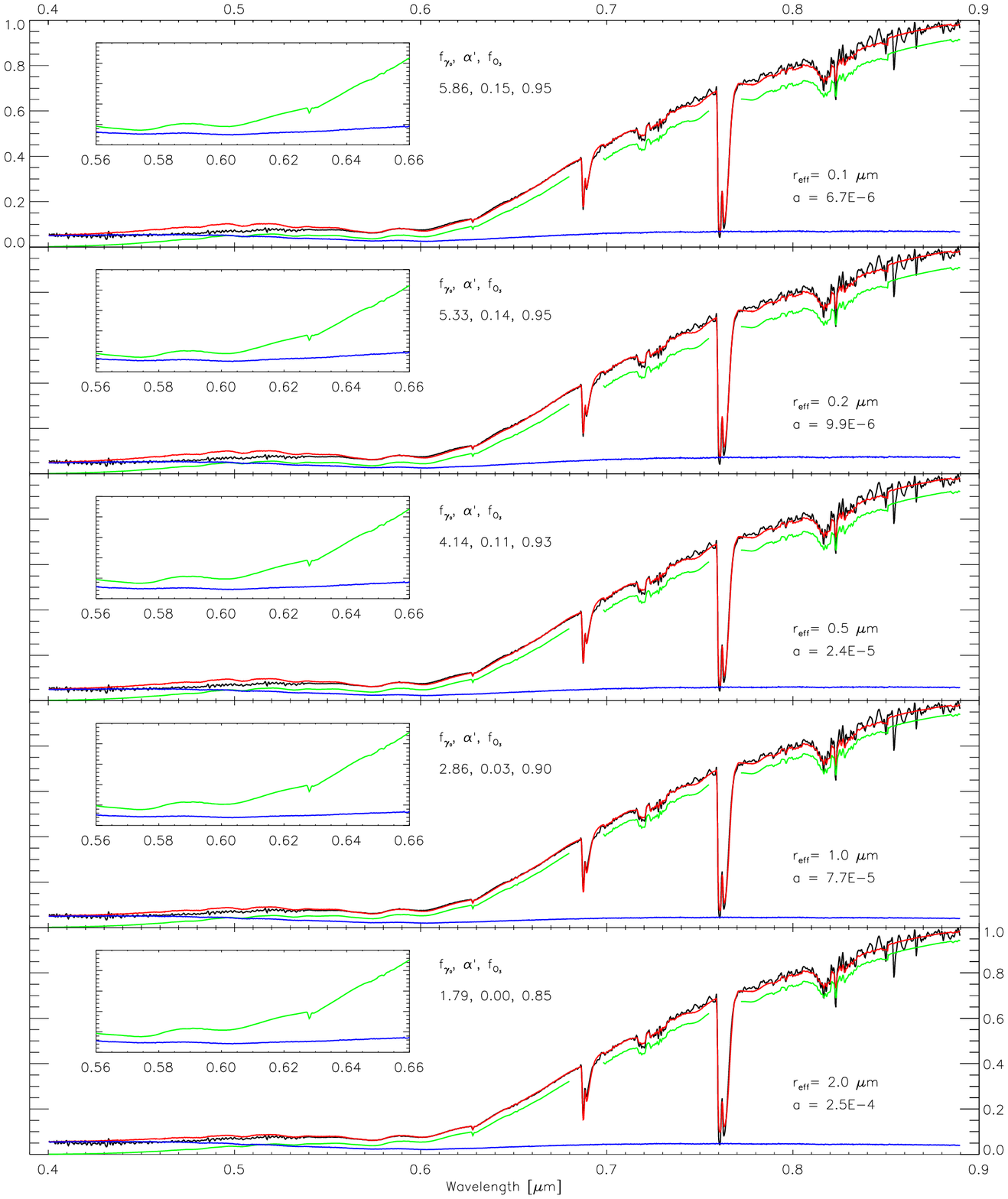}
\end{figure}

\end{document}